\documentclass[aps,prl,twocolumn,showpacs,10pt,superscriptaddress]{revtex4-1}
\usepackage{graphicx,amsmath,amssymb,amsfonts,sidecap,color}

\usepackage{dsfont}

\usepackage{color}

\newcommand{\be}{\begin{equation}}
\newcommand{\ee}{\end{equation}}

\begin{document}

\title{Viewing Majorana Bound States by Rabi Oscillations}

\author{Zhi Wang}
\affiliation{School of Physics and Engineering, Sun Yat-sen University, Guangzhou 510275, China}
\affiliation{International Center for Materials Nanoarchitectonics (WPI-MANA)\\
National Institute for Materials Science, Tsukuba 305-0044, Japan}
\author{Qi-Feng Liang}
\affiliation{Department of Physics, Shaoxing University, Shaoxing 312000, China}
\affiliation{International Center for Materials Nanoarchitectonics (WPI-MANA)\\
National Institute for Materials Science, Tsukuba 305-0044, Japan}

\author{Dao-Xin Yao}
\affiliation{School of Physics and Engineering, Sun Yat-sen University, Guangzhou 510275, China}

\author{Xiao Hu}
\affiliation{International Center for Materials Nanoarchitectonics (WPI-MANA)\\
National Institute for Materials Science, Tsukuba 305-0044, Japan}

\begin{abstract}
Rabi oscillation is a pure quantum phenomenon where the system jumps forth and back between two quantum levels under stimulation of
a microwave, and a resonance occurs when the energy difference is matched by the photon energy.
Rabi oscillations have been observed in various quantum systems so far, and especially are used to demonstrate the quantum coherence of quantum bits.
In the present work, we explore Rabi oscillation in a system accommodating the elusive Majorana bound states (MBSs) under intensive search recently.
The proposed setup is constructed by a quantum dot (QD) and a superconducting quantum interference device (SQUID), where a semiconductor nanowire with spin-orbital coupling in the topological state is introduced to form Josephson junction. When the coupling between QD and the Josephson junction is tuned by an ac gate voltage, Rabi oscillation takes place among quantum states formed by QD and MBSs, which makes it possible to reveal bizarre properties of MBSs by sensing the electron occupation on QD. Especially, one should be able to observe
the fractional Josephson relation $\cos(\pi\Phi/\Phi_0)$ unique to MBSs with $\Phi$ the magnetic flux applied in the SQUID and $\Phi_0=hc/2e$ the flux quantum. The system has been investigated
in terms of the analytic Floquet theorem and numerical simulations with fine agreement.
\end{abstract}

\date{\today}

\maketitle

Majorana bound states (MBSs) in topological superconductor become the focus of many researches in recent years\cite{read,kitaev,kanermp,alicearpp,beenakkerarcp}. These bizarre quasi-particles are equal quantum superposition of electrons and holes, and obey the non-Abelian statistics\cite{ivanov,sarmarmp,sternscience}. Pairs of MBSs constitute topological quantum bits (qubits), and information processing can be performed by braiding MBSs\cite{ivanov,aliceanphy}. Decoherence suffered by many other approaches to realizing qubits is expected to be suppressed by exploiting the nonlocal and charge-neutral
property of MBSs\cite{kanermp,jiang,bonderson,pekker}.

MBSs have been theoretically predicted in several spin-orbit-coupled superconducting systems, including superconductor-topological insulator interfaces\cite{kanefu}, semiconductor-superconductor heterostructures\cite{sarma1d,sarma2d,oppen1d}, and nanotube-superconductor devices\cite{loss12}. In particular, a nanowire of spin-orbit-coupled semiconductor in proximity to a conventional superconductor is promising \cite{sarma1d,oppen1d}, which is expected to support MBSs at the two ends under appropriate Zeeman energy and chemical potential.

Exploring physical properties unique to MBSs is very important \cite{sarma12,Ioselevich,jiang11,lawprb,lawarxiv,law12,loss13}, since on one hand they can distinguish MBSs from other accidental zero-energy Andreev bound states, and on the other hand they derive novel quantum functionalities useful for future applications.
Unconventional transport properties of MBSs, such as the resonant Andreev reflection\cite{lawrar}, the crossed Andreev reflection\cite{beenakkercar}, and the fractional Josepshon effect\cite{kitaev,sarma1d} have been investigated in recent experiments \cite{kouwenhoven,rokhinson,heiblum,xu,rodrigo,harlingen} with promising signatures, which definitely have generated huge excitements.

\begin{figure}[b]
\begin{center}
\includegraphics[clip=true,width=0.9\columnwidth]{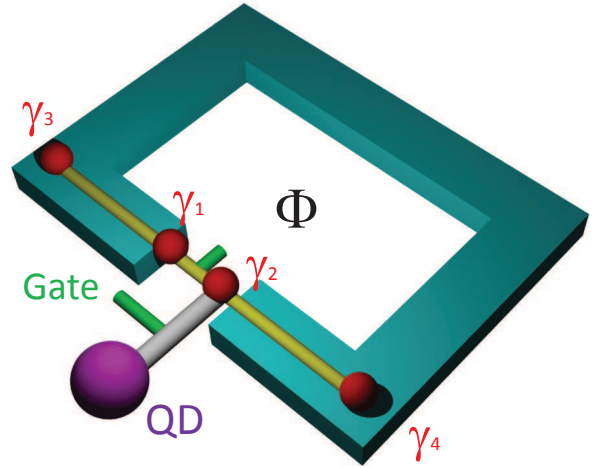}
\caption{(Color online). Schematic setup of a hybrid device constructed by a topological rf-SQUID and a quantum dot.
The quantum dot is tunneling connected to one Majorana bound state in the SQUID, with the tunneling strength tunable by an
ac gate voltage. }
\end{center}
\end{figure}

In this work, we discuss theoretically Rabi oscillations associated with MBSs. Having been observed in many systems, Rabi oscillations take place between two quantum levels when the frequency of an ac driving matches the energy difference. Observation on Rabi oscillation is taken as the hallmark of successful construction of qubits\cite{nakamura,mooij,yamamoto,bruder,tarucha}, which provide important information on the quantum states without violating the quantum coherence. Exploring Rabi oscillations associated with MBSs is expected to be able to provide important features unique to MBSs by illuminating directly the detailed energy levels,
an approach lacking up to the moment of this writing. Especially, we propose a setup as shown in Fig.~1, which is constructed by a Majorana rf-SQUID and a nearby quantum dot (QD). The time evolution of
the system is investigated by the Floquet theorem and numerical technique. Intriguingly we find that the driving frequency for resonant Rabi oscillation is modulated by the magnetic flux through the SQUID in the
form of fractional Josephson relation $\cos(\pi\Phi/\Phi_0)$, which is unique to MBSs. This behavior can be observed experimentally by sensing the charge on QD with well established techniques.

\begin{figure}[t]
\begin{center}
\includegraphics[clip=true,width=1\columnwidth]{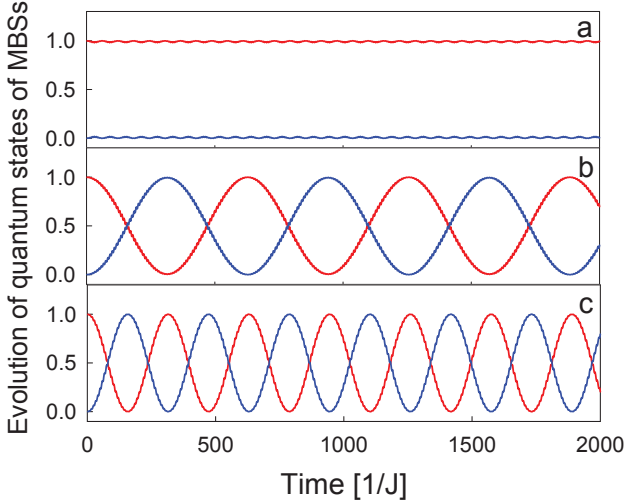}
\caption{(Color online). Time evolution of the weights $|P_1(t)|^2$ (red) and $|P_2(t)|^2$ (blue) for the two quantum states with the initial condition $P_1(0) = 1$, (a) for off-resonance with driving frequency $\Omega=2.3J$ and $T_1= 0.01J$, (b) and (c) for on-resonance with $\Omega=2.402J$ and $T_1= 0.01J$ and $T_1=0.02J$ respectively. Other parameters are $T_0 = 0.05J$, $\delta_{\rm L} = 0.02J$, $\delta_{\rm R}=0.005J$, $\Phi=0$ and $\epsilon=0.4J$. $\hbar=1$ is taken through the paper.
}
\end{center}
\end{figure}

\section{\uppercase\expandafter{\romannumeral1}. DEVICE WITH MBSs}

The Majorana SQUID is constructed by a semiconductor nanowire with spin-orbit coupling put on a conventional s-wave superconductor with ring shape (see Fig.~1).
When the chemical potential and Zeeman field are tuned appropriately, the semiconductor nanowire enters the topological superconducting phase due to proximity effect\cite{kouwenhoven}. A tunneling barrier is introduced at the center of the nanowire for example by a dc voltage, resulting in a Josephson junction. The Hamiltonian for low-energy physics of the Majorana SQUID is given by\cite{kitaev,aliceanphy}

\begin{eqnarray}
\mathcal{H}_{M} = i J \cos \frac{\phi}{2} \gamma_1 \gamma_2  + i\delta_{\rm L} \gamma_1 \gamma_3 + i\delta_{\rm R} \gamma_4 \gamma_2,
\end{eqnarray}
where the first term is the fractional Josephson energy with $J$ the energy integral of the junction, $\phi=2\pi \Phi/ \Phi_0$ the phase difference across the junction induced by the magnetic flux $\Phi$ through the SQUID with $\Phi_0$ the flux quantum;
$\gamma_1$ and $\gamma_2$ are the two MBSs at the two sides of the junction while $\gamma_3$ and $\gamma_4$ are the two MBSs at the two ends of the wire, as illustrated in Fig.~1; $\delta_{\rm L}$ and $\delta_{\rm R}$ are couplings among MBSs due to small wave-function overlaps in the two segments.

The QD is prepared in such a way that only one level is energetically relevant, and thus its state is described by
$\mathcal{H}_{D} = \epsilon d^{\dagger} d$
with $d^{\dagger}$ the electron creation operator and $\epsilon$ the electron occupation energy.
The QD is tunneling connected to one of the MBSs at the Josephson junction with $\mathcal{H}_T = i T (d^\dagger+d) \gamma_2$,
where $T$ is the coupling strength controlled by an ac gate voltage (see Fig.~1) and changes with time periodically
$T = T_0 + 2T_1 \cos \Omega t$.

The low-energy dynamics of the present device is described by the time-dependent Schr\"{o}dinger equation,
\begin{equation}
i \frac {d} {dt} |G\rangle= \mathcal{H} |G \rangle=(\mathcal{H}_{M} +\mathcal{H}_{D} + \mathcal{H}_T) |G \rangle,
\end{equation}
where $|G \rangle$ is the quantum state of the system within the eight-dimensional Hilbert space spanned by the parity states of MBSs and the electron occupation of QD, and $\hbar=1$ is taken. Defining two fermionic operators with the four MBSs
$f_1^{\dagger} =(\gamma_1 - i \gamma_2)/2$ and $f_2^{\dagger} =(\gamma_4 - i \gamma_3)/2$,
the basis functions for Hamiltonian $\mathcal{H}$ can be set as
$|0\rangle(\equiv|S_1\rangle)$, $f_1^\dagger d^\dagger |0\rangle(\equiv|S_2\rangle)$, $f_2^\dagger f_1^\dagger |0\rangle(\equiv|S_3\rangle)$, $f_2^\dagger d^\dagger |0\rangle(\equiv|S_4\rangle)$, $f_1^\dagger |0\rangle$, $d^\dagger |0\rangle$, $f_2^\dagger|0\rangle$, $f_2^\dagger f_1^\dagger d^\dagger |0\rangle$,
where $|0\rangle$ is the vacuum for operators $f_1$, $f_2$ and $d$. In this basis, the Hamiltonian $\mathcal{H}$ is an 8$\times$8 matrix.
With the conservation of total parity upon application of gate voltage and Cooper pair tunneling, Hamiltonian $\mathcal{H}$ is block-diagonal. Without losing generality, we hereafter focus on the  even-parity subspace of the system. The Schr\"{o}dinger equation (2) then reads
\begin{widetext}
 \begin{eqnarray}\label{eq1}
  {i  } \frac {d} {dt} \left(\begin{array}{cccc}  P_1  \\\ P_2 \\\  P_3   \\\  P_4  \end{array} \right)
     =
  \left(\begin{array}{cccc}
 -{J}\cos \frac{\phi}{2}& T_0 + 2T_1 \cos \Omega t                 &\delta_{\rm L}-\delta_{\rm R}&0\\
 T_0 + 2T_1 \cos \Omega t &  {J}\cos \frac{\phi}{2}+\epsilon     &0&\delta_{\rm L}+\delta_{\rm R}\\
\delta_{\rm L}-\delta_{\rm R}&0&               {J}\cos \frac{\phi}{2} &T_0 + 2T_1 \cos \Omega t \\
0&\delta_{\rm L}+\delta_{\rm R}&       T_0 + 2T_1 \cos \Omega t&-{J}\cos\frac{\phi}{2} +\epsilon \\
   \end{array}\right)
     \left(\begin{array}{cccc}  P_1  \\\ P_2   \\\  P_3 \\\  P_4  \end{array} \right),
\end{eqnarray}
\end{widetext}
with $|G\rangle=\sum_j P_j|S_j\rangle$.
It is noted that the parameters in the present system are highly controllable, with $\Phi$ by the applied flux through the SQUID and $T$ by the gate voltage, which makes
a comprehensive exploration of MBSs in terms of Rabi oscillations possible.

\section{\uppercase\expandafter{\romannumeral2}. RABI OSCILLATIONS AND FLOQUET THEORY}

Typical time evolutions of the quantum state obtained by numerical integration of Eq.~(3) are given in Fig.~2 for
an initial state at $|S_1\rangle$. When the driving frequency is off resonance, the system stays at the initial state, while for the driving frequency matching the energy difference between $|S_1\rangle$ and $|S_2\rangle$, the system oscillates between the two states as $|P_1|^2(t) =\cos^2(T_1 t)$ and $|P_2|^2(t) =\sin^2 (T_1 t)$.
In Fig.~3 we map out the full spectrum of resonant driving frequency for Rabi oscillations by sweeping the magnetic flux in the SQUID, where one observes two sets of curves, four vertical lines and one horizontal line.

In order to understand these rich Rabi oscillations, we analyze the system in terms of the Floquet theorem\cite{shirley}. Since the system is driven by the ac voltage, the solution of
Eq.~(3) should be given in the form $ \Psi(t) \equiv (P_1, P_2, P_3, P_4)^{\rm t} = \psi(t) e^{-iQt}$, with $Q$ to be determined and $\psi(t)$ a periodic function $\psi(t)=\psi(t+\frac{2\pi}{\Omega})$. The Fourier components of the wave-function $\psi(t)$ and the Hamiltonian are given as
\begin{eqnarray}
\psi(t) = \sum_n \psi_n e^{i n \Omega t},  \hspace{5mm}  {\mathcal{H}} = \sum_n {\mathcal{H}}_n e^{i n \Omega t},
\end{eqnarray}
with
 \begin{eqnarray}
{\mathcal{H}}_0 =  \left(\begin{array}{cccc}
 -{J}\cos \frac{\phi}{2}& T_0                 &\delta_{\rm L}-\delta_{\rm R}&0\\
 T_0  &  {J}\cos \frac{\phi}{2}+\epsilon     &0&\delta_{\rm L}+\delta_{\rm R}\\
\delta_{\rm L}-\delta_{\rm R}&0&               {J}\cos \frac{\phi}{2} &T_0 \\
0&\delta_{\rm L}+\delta_{\rm R}&       T_0 &-{J}\cos\frac{\phi}{2} +\epsilon \\
   \end{array}\right),\nonumber\\
\end{eqnarray}
\vspace{-8mm}
 \begin{eqnarray}
{\mathcal{H}}_1 = {\mathcal{H}}_{-1}=  \left(\begin{array}{cccc}
 0& T_1 &0&0\\
 T_1 & 0&0&0\\
0&0& 0 & T_1  \\
0&0& T_1&0\\
   \end{array}\right),
\end{eqnarray}
and all other matrices zero in the present case. Plugging them into Eq.~(3), we arrive at the following static secular equation\cite{shirley},
\begin{eqnarray}
\sum_{l} \left({\mathcal{H}}_{n-l} + n\Omega \delta_{nl} \hat I\right) \psi_l = Q \psi_n,
\end{eqnarray}
with $\hat I$ the 4$\times$4 identity matrix.
Defining a vector $ \tilde \psi = (..., \psi^{\rm t}_{-1}, \psi^{\rm t}_0, \psi^{\rm t}_1,...)^{\rm t}$, we obtain the time-independent Floquet Hamiltonian
\begin{eqnarray}
{\mathcal{H}}_{\rm f} =
\left(\begin{array}{ccccccc}
 \cdot&\cdot&\cdot&\cdot&\cdot&\cdot&\cdot\\
\cdot& {\mathcal{H}}_0-2\Omega \hat I & {\mathcal{H}}_1&0& 0&0& \cdot\\
\cdot& {\mathcal{H}}_1 & {\mathcal{H}}_0-\Omega \hat I&{\mathcal{H}}_1& 0&0& \cdot\\
\cdot& 0 & {\mathcal{H}}_1& {\mathcal{H}}_0& {\mathcal{H}}_1&0& \cdot\\
\cdot& 0 & 0&{\mathcal{H}}_1& {\mathcal{H}}_0+\Omega \hat I&{\mathcal{H}}_1& \cdot\\
\cdot& 0 & 0&0& {\mathcal{H}}_1&{\mathcal{H}}_0+2\Omega \hat I& \cdot\\
 \cdot&\cdot&\cdot&\cdot&\cdot&\cdot&\cdot\\
   \end{array}\right).\nonumber\\
\end{eqnarray}

\begin{figure}[t]
\begin{center}
\includegraphics[clip=true,width=1\columnwidth]{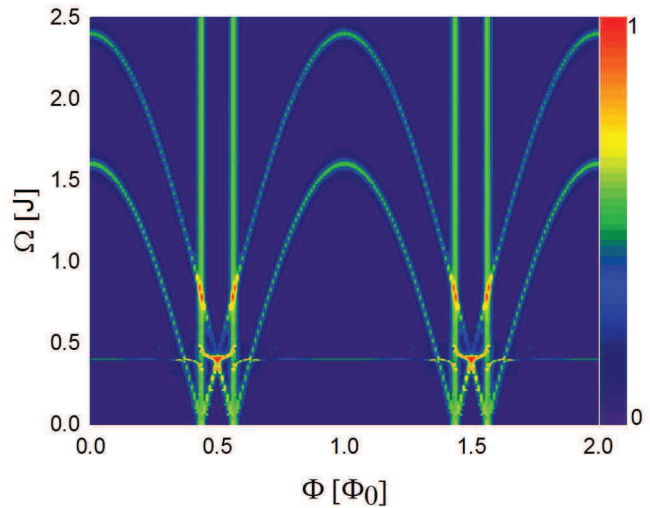}
\caption{(Color online). Driving frequency $\Omega$ for coherent Rabi oscillation as a function of applied flux $\Phi$. The color is for the strength of Rabi oscillation measured by the peak-dip difference in the oscillation of the occupation probability. The initial state is presumed as an equal weight of $|S_1\rangle$ and $|S_3\rangle$ associated with empty QD. Parameters are the same as Fig.~2 except for $T_1=0.015J$.}
\end{center}
\end{figure}

Now the original system of four states with time-periodic Hamiltonian ${\mathcal{H}}$ in Eq.~(3) is transformed into a system with static Floquet Hamiltonian ${\mathcal{H}}_{\rm f}$.
The diagonal part of the Floquet Hamiltonian is built by infinite $4\times 4$ blocks, with each block formed by the time-invariant part of the original Hamiltonian ${\mathcal{H}}_0$ adding energy quanta $n\Omega$
associated with integer number of "photons"; the diagonal blocks with one-photon difference are connected by an off-diagonal 4$\times$4 block given by the amplitude of time-periodic part in the
original Hamiltonian. The basis states for the Floquet Hamiltonian ${\mathcal{H}}_{\rm f}$ are the Floquet states $|n,\alpha \rangle$, with $\alpha$ referring to the four states $|S_j\rangle$ ($j=1$, 2, 3 and 4)
and $n$ to the Fourier component.

Within the Floquet theory\cite{shirley}, it has been shown that the transition probability between two quantum states $|\alpha\rangle$ and $|\beta\rangle$
can be expressed as a summation of those between corresponding Floquet states $|0,\alpha\rangle$ and $|n, \beta\rangle$
\be
T_{|\alpha \rangle \rightarrow |\beta\rangle} = \sum_{n} \left|\langle n,\beta|e^{-i{\mathcal{H}}_f (t-t_0)}|0,\alpha\rangle \right|^2.
\ee
Therefore, the problem of evaluating the transition probability between two quantum states governed by time-dependent Hamiltonian is reduced to a corresponding one
with time-independent Floquet Hamiltonian, with the latter being a conventional problem in quantum mechanics.

For example, let us consider the probability of the transition between quantum states $|S_1\rangle$ and $|S_2 \rangle$.
With a second-order perturbation treatment, we obtain the effective 2$\times$2 Hamiltonian (see Supplement) for transition between
Floquet states $|0,S_1\rangle$ and $|-1,S_2\rangle$
\begin{equation}
{\mathcal{H}}_{e} = \left(\begin{array}{cc}
  -{J}\cos \frac{\phi}{2} + E_{-} &T_1 \\
    T_1   &  {J}\cos \frac{\phi}{2}  +\epsilon - \Omega + E_{+}
   \end{array}\right),
\end{equation}
with energy shift $E_{\pm}= \pm(T_0^2+T_1^2/2)/(2J \cos\frac{\phi}{2}+ \epsilon) \pm(\delta_{\rm L}\pm \delta_{\rm R})^2/J\cos\frac{\phi}{2}$.
Starting from $|S_1\rangle$, the system evolves with time according to
\be
|P_1|^2(t) =\frac{T_1^2}{\omega^2 }\cos^2 (\omega t), \hspace{5mm} |P_2|^2(t) =\frac{T_1^2}{\omega^2 }\sin^2 (\omega t),
\ee
with $\omega=\sqrt{T_1^2+\left[\Omega- 2J\cos(\phi/2)-{\epsilon} - E_1 \right]^2}$ and
$E_1=E_{+}-E_{-}=(2T_0^2+T_1^2)/(2J \cos\frac{\phi}{2}+\epsilon)+2(\delta^2_{\rm L}+\delta_{\rm R}^2)/J\cos\frac{\phi}{2}$.
When the photon energy of the driving ac voltage fills the energy gap, a coherent Rabi oscillation between the two levels
appears characterized by the maximal oscillation amplitude in the occupation probability.
In the same way, one has the contribution from the Floquet state $|1,S_2 \rangle$. The
spectrum for coherent Rabi oscillation among $|S1 \rangle$ and $|S2 \rangle$ is then given by
\begin{eqnarray}
\Omega_1= |2J\cos ({\phi}/{2})+\epsilon+ E_1|.
 \end{eqnarray}
With the same procedure, we can obtain the Rabi oscillation between $|S_3\rangle$ and $|S_4\rangle$ under the driving frequency
 \begin{eqnarray}
\Omega_2= |2J\cos ({\phi}/{2})-\epsilon+ E_2|,
 \end{eqnarray}
with $E_2= (2T_0^2+T_1^2)/(2J \cos\frac{\phi}{2}- \epsilon)+2(\delta^2_{\rm L}+\delta_{\rm R}^2)/J\cos\frac{\phi}{2}$.

Equations (12) and (13) are the main results of this work, which explain the two sets of curves in Fig.~3.
We emphasize that the driving frequencies $\Omega_1$ and $\Omega_2$ for resonating Rabi oscillations vary with the applied magnetic flux in the way of $\cos(\pi\Phi/\Phi_0$), characterizing the fractional Josephson effect, which is an intrinsic feature of MBSs.
Since the electron occupation on QD in state $|S_1\rangle$ (or $|S_3\rangle$) is different from that in state $|S_2\rangle$ (or $|S_4\rangle$),
the Rabi oscillations can be detected by measuring the charge on QD, which can be carried out experimentally via well established techniques\cite{petersson}.

\section{\uppercase\expandafter{\romannumeral3}. MANY-BODY ENTANGLEMENT}

When the denominators in energy shifts $E_{\pm}$ (and thus $E_{1,2}$) become zero, the simple treatment in the Floquet theorem based on
$2\times 2$ matrix breaks down. The four vertical lines in Fig.~3 are associated with $\epsilon=\pm 2J\cos(\phi/2)$, where
$|S_1\rangle$ and $|S_2\rangle$ or $|S_3\rangle$ and $|S_4\rangle$ are degenerate in energy irrespective of driving frequency,
and fall into Rabi oscillation respectively.
When the curves given by Eqs.~(12) and (13) cross with the vertical lines (see Fig.~3), two sets of coherent Rabi oscillations, $|S_1\rangle \leftrightarrow |S_2\rangle$ and $|S_3\rangle \leftrightarrow |S_4\rangle$ take place
simultaneously.

There is a horizontal line at $\Omega\simeq \epsilon$ in Fig.~3, at which two Rabi oscillations $|S_1\rangle \leftrightarrow |S_4\rangle$ and $|S_2\rangle \leftrightarrow |S_3\rangle$ take place simultaneously.  One notices that
in Eqs.~(5), (6) and (8), there is no matrix element between the Floquet states $|0,S_1\rangle$ and $|-1,S_4\rangle$.
Instead, they are connected via intermediate states $|-1,S_2\rangle$ and $|0,S_3\rangle$. In terms of second-order
perturbation, one can evaluate the coupling element as
$T_1\delta_{\rm L}/J\cos(\phi/2)$ and corresponding energy shifts (see Supplement). Similar results can be obtained for
$|S_2\rangle$ and $|S_3\rangle$. The frequencies of these Rabi oscillations are smaller than those discussed above by an order of
$\delta_{\rm L}/J$ since higher-order processes are involved. Only $\delta_{\rm L}$ appears in the
Rabi frequency, due to the structure of setup where QD is connected to the right MBS at the junction (see Supplement).

Interesting physics takes place at the point $\Phi=\Phi_0/2$ and $\Omega\simeq \epsilon$, where the horizontal line and the two cosine curves cross in Fig.~3.
Since, $|0,S_1\rangle$, $|-1,S_2\rangle$, $|0,S_3\rangle$ and $|-1,S_4\rangle$ take the same energy at this point, the four states $|S_1\rangle$, $|S_2\rangle$, $|S_3\rangle$ and $|S_4\rangle$ entangle with each other,
as shown in Fig.~4.
At $\Phi= \Phi_0 /2$, there is no coupling among the two MBSs $\gamma_1$ and $\gamma_2$. It is
then better to describe the system by parity of electron number on the left segment formed by $\gamma_1$ and $\gamma_3$, and that on the right segment
formed by $\gamma_2$ and $\gamma_4$. It is not difficult to see that the "bonding" state
$|S^+_{13}\rangle =(|0,S_1\rangle+|0,S_3\rangle)/\sqrt{2}$ corresponds to even-parity states on both left and right segments, while
$|S^+_{24}\rangle =(|-1,S_2\rangle+|-1,S_4\rangle)/\sqrt{2}$ corresponds to even-parity state on the left segment while odd-parity state on the right segment. The two states
are coupled via $T_1$ which changes the parity of the right segment by moving one electron from or to QD. The same physics applies for the two anti-bonding states
$|S^-_{13}\rangle =(|0,S_1\rangle-|0,S_3\rangle)/\sqrt{2}$ and $|S^-_{24}\rangle =(|-1,S_2\rangle-|-1,S_4\rangle)/\sqrt{2}$, except for that the left segment is in odd-parity state.

The bonding and antibonding states are not coupled directly since QD is attached only to the right segment in the present setup. Therefore, the same treatment based on $2\times 2$ matrix
can be applied to the bonding and antibonding blocks separately. No fully resonant Rabi oscillation can
take place at this specific point of driving frequency and applied flux, which gives rise two Rabi oscillations with different frequencies (see Eq.~(11) and Supplement), although the off-diagonal elements for the two blocks are both $T_1$.
Starting from $|S_1\rangle$,
the system is in a superposition of even- and odd-parity states in the left segment, and thus evolves with time with two Rabi frequencies, which results in the beating waves in QD occupation probability as shown in Fig.~4.

\begin{figure}[t]
\begin{center}
\includegraphics[clip=true,width=1\columnwidth]{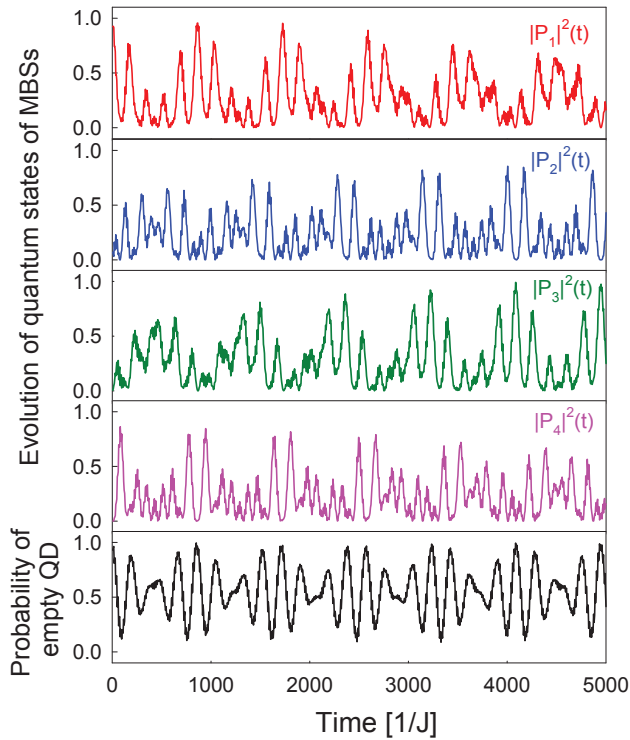}
\caption{(Color online). Oscillations among the four states and charge occupation on QD at $\Phi=\Phi_0/2$ and $\Omega= \epsilon$, with $P_1(0)=1$.
}
\end{center}
\end{figure}

\section{\uppercase\expandafter{\romannumeral4}. DISSCUSSIONS}

The interactions between MBSs in each of the two segments have been taken into account in the present study. Neglecting
these interactions, an approximation justified only in a sufficiently long nanowire, the Hamiltonian (3) is reduced to
two block-diagonalized 2$\times$2 matrices reflecting the fact that parity of subsystem consisted of junction qubit and QD is preserved.
In this case, only one cosine curve in Fig.~3 can be observed associated with Rabi oscillation either
$|S_1\rangle \leftrightarrow |S_2\rangle$ or $|S_3\rangle \leftrightarrow |S_4\rangle$.
The driving frequency for resonant Rabi oscillation then exhibits a period of 2$\Phi_0$ with respect to the applied magnetic flux,
which is a manifestation of the $4\pi$-period current-phase relation associated with MBSs discussed in literature
\cite{kitaev,aliceanphy}. As far as interactions among MBSs in the two segments are finite, the period of
driving frequency for resonant Rabi oscillation with respect to magnetic flux is $\Phi_0$ (or equivalently 2$\pi$ in phase) as shown in Fig.~3,
where two cosine curves exist and cross each other at $\Phi_0/2$ in a symmetric way. It is noticed that even in this case the
fractional Josephson relation featured by the MBSs is clearly detectable by Rabi oscillation from the detailed magnetic-flux dependence of the
two cosine curves as given explicitly in Eqs.~(12) and (13). This is the advantage of the present approach based on Rabi oscillation, which
accesses directly the energy levels composed by MBSs.

We have concentrated on the even-parity subspace of the system in Eq.~(3). It is easy to see that the
physics for the odd-parity subspace is the same, since formally the two subspaces can be transformed
to each other by redefining the sign of interactions $J$ and $\delta_{\rm R}$.

Let us discuss the experimental relevance of our results. The device under consideration makes use of one-dimensional
topological superconductor and a QD. Topological superconductivity and MBSs as its novel quasiparticle
excitations have been reported recently in a system of nanowire of spin-orbit-coupled semiconductor in proximity to a conventional superconductor with the state-of-the-art technology\cite{kouwenhoven}. The Rabi oscillations revealed in the present work can be measured by sensing the charge on QD, which is a well-established experimental technique by now, thanking to the research for QD qubit in recent years\cite{petersson}. Therefore, the present proposal should be experimentally accessible. Especially, we wish to notice that all physical quantities important to low-energy physics of the system can be detected experimentally through the resonating curves shown in Fig.~3, which provides detailed information on the quantum property of MBSs.

We can make an estimate on the frequency to drive coherent Rabi oscillations in the present system. The typical proximity induced superconducting gap in the semiconductor nanowire is around $\Delta\simeq 200\mu$eV \cite{kouwenhoven}, which sets the largest energy scale; the coupling energy at Josephson junction is controlled to be smaller than $\Delta$ by one order in magnitude ($J\simeq 20\mu$eV), and all other energies should also be small as compared with the energy gap $\Delta$. With the energy parameters investigated in the present work, the driving frequencies for resonant Rabi oscillation are in the range of GHz. In order to provide topological protection to the MBSs, temperature is to be kept below $\sim 2$ Kelvin.

In summary, we propose a hybrid device constructed by a Majorana rf-SQUID and a quantum dot.
We show that the ac voltage on gate between SQUID and quantum dot can drive coherent Rabi oscillations among
quantum states built by Majorana bound states, which can be measured by sensing the charge on the quantum dot. Associated with the Majorana bound states, the frequencies of Rabi oscillations vary with the applied magnetic flux through the rf-SQUID in the form of fractional Josephson relation $\cos(\pi\Phi/\Phi_0)$, with $\Phi_0$ the flux quantum.
This unique feature is expected to be useful for identifying and manipulating Majorana bound states.

\section{\uppercase\expandafter{\romannumeral6}. SUPPLEMENTARY MATERIAL}

\begin{center}
\noindent\textbf{A. Model}\\
\end{center}

Here we show the details for building the minimum model to describe our setup. The SQUID of the setup is build by a circled conventional superconductor with a straight spin-orbit coupling nanowire on top of it, as shown in Fig.~1 of the main text. The proximity effect induces a gap in the wire, making it superconducting.
Under an appropriate magnetic field, the nanowire enters into a topological state and becomes a topological superconductor. There are two domain walls between the topological superconducting wire and the trivial superconducting base. A gate voltage is applied in the center of the wire, producing a well-defined tunneling junction. For convenience we call the two parts of nanowire as left and right segments. The critical current of this junction is small enough, so that the superconducting phase change across this junction is equal to the magnetic flux through the SQUID. Meanwhile, the nanowire is much smaller than the base, that the superconducting phase on the wire can be safely assumed to be the same within the segments. The low-energy physics the system can be described by four
Mjorana bound states (MBSs), two locating at the junction, and two at the domain walls.
From the work of Kitaev\cite{kitaev}, the MBSs at the junction are defined as

\begin{eqnarray}
\gamma_1 &&= e^{i\frac{\theta_L}{2}} \hat c_1 + e^{-i\frac{\theta_L}{2}} \hat c^\dagger_1\nonumber \\\
\gamma_2 &&= -ie^{i\frac{\theta_R}{2}} \hat c_2 + ie^{-i\frac{\theta_R}{2}} \hat c^\dagger_2
\end{eqnarray}
where $\theta_{L,R}$ are the superconducting phase of the left/right segment, $\hat c_{1,2}$ are the local fermion operators at the two sides of the junction. Similarly, the other two MBSs on the domains are defined as
\begin{eqnarray}
\gamma_3 &&= -i e^{i\frac{\theta_L}{2}} \hat c_3 + i e^{-i\frac{\theta_L}{2}} \hat c^\dagger_3\nonumber \\\
\gamma_4 &&= e^{i\frac{\theta_R}{4}} \hat c_2 + e^{-i\frac{\theta_R}{2}} \hat c^\dagger_4
\end{eqnarray}
where $\hat c_{3,4}$ are the local fermion operators at the domain walls. For each of the four MBSs, there is a partner, with which
the local fermion operator is reconstructed as
\begin{eqnarray}
\hat c_1 &&= e^{-i\frac{\theta_L}{2}} (\gamma_1 +i\gamma'_1)/2     \nonumber \\\
\hat c_2 &&= e^{-i\frac{\theta_R}{2}} (\gamma'_2 +i\gamma_2)/2     \nonumber \\\
\hat c_3 &&= e^{-i\frac{\theta_L}{2}} (\gamma'_3 +i\gamma_3)/2     \nonumber \\\
\hat c_4 &&= e^{-i\frac{\theta_R}{2}} (\gamma_4 +i\gamma'_4)/2.
\end{eqnarray}
It is noted that $\gamma'_{1,2,3,4}$ correspond to excitations above the superconducting energy gap, and thus are irrelevant to low-energy physics.

The two MBSs in each segment have a small coupling due to the finite wire length, resulting in the coupling Hamiltonian given first by Kitaev\cite{kitaev}
\begin{eqnarray}
\mathcal{H}_{\delta} = i \delta_L \gamma_1 \gamma_3 + i \delta_R \gamma_4 \gamma_2.
\end{eqnarray}
The coupling energy $\delta_{L,R}$ is considered as a small parameter in the present work.
The two segments are coupled at the junction through the electron tunneling
\begin{eqnarray}
\mathcal{H}_{J} && = {J} (\hat c_1^\dagger \hat c_2 + \hat c_2^\dagger \hat c_1) \nonumber\\\
&& = {J}  e^{\frac{i\phi}{2}} (\gamma_1 \gamma'_2 -i\gamma'_1 \gamma'_2 +i\gamma_1\gamma_2 +\gamma'_1 \gamma_2)/4   \nonumber\\\
 &&+ {J} e^{-\frac{i\phi}{2}} (\gamma'_2 \gamma_1 - i\gamma_2 \gamma_1 +i\gamma'_2\gamma'_1 +\gamma_2 \gamma'_1)/4
 \nonumber\\\
 && \approx iJ \cos \frac{\phi}{2} \gamma_1 \gamma_2,
\end{eqnarray}
with $\phi=\theta_L-\theta_R$ the phase difference across the junction, where all excitations above the superconducting gap are neglected presuming that the
system is kept at sufficiently low temperature. This fractional Josephson energy was derived by Alicea \textit{et al}\cite{aliceanphy}.
These two energies are contributed only from MBSs, and are summarized in Eq.~(1) in the main text

\begin{widetext}
\begin{eqnarray}
\mathcal{H}_{M} = \mathcal{H}_{\delta} + \mathcal{H}_{J}.
\end{eqnarray}

The quantum dot (QD) is prepared in a way that there is only one energy level:
\begin{eqnarray}
\mathcal{H}_{D} = \epsilon d^\dagger d,
\end{eqnarray}
where $d^\dagger$ is the electron creation operator and $\epsilon$ is the occupation energy.
The QD is connected to the right side of the junction in a tunneling way with the tunneling Hamiltonian
\begin{eqnarray}
\mathcal{H}_{T} && = 2T (\hat c_2^\dagger  d +  d^\dagger \hat c_2) \nonumber\\\
&& = T \left[ e^{\frac{i\theta_R}{2}} (\gamma'_2 -i\gamma_2)  d  + d^\dagger  e^{-\frac{i\theta_R}{2}} (\gamma'_2 +i\gamma_2) \right]
\nonumber\\\
 && \approx T\left[  -i e^{\frac{i\theta_R}{2}} \gamma_2 d +  i e^{-\frac{i\theta_R}{2}}   d^\dagger \gamma_2     \right]
\nonumber\\\
&&=iT (d^\dagger +d ) \gamma_2,
\end{eqnarray}
where high-energy excitations are neglected as above, and $\theta_R = 0$ is taken since superconducting phase itself can be chosen arbitrarily as long as the phase difference remains the same.

Now we formulate the total Hamiltonian into the most convenient matrix form. For this purpose, we redefine fermionic operators,
\begin{eqnarray}
f^\dagger_1 = (\gamma_1 - i \gamma_2)/2, f^\dagger_2 = (\gamma_4 - i \gamma_3)/2
\end{eqnarray}
The MBS operators can be rewritten as
\begin{eqnarray}
\gamma_1 = f_1^\dagger +f_1,
\gamma_2 = i(f_1^\dagger - f_1),
\gamma_3 = i (f_2^\dagger -f_2),
\gamma_4 = f_2^\dagger + f_2
\end{eqnarray}
and the Hamiltonian is given by

\begin{eqnarray}
\mathcal{H}_{M} = J \cos \frac{\phi}{2} (2 f^\dagger_1 f_1 - 1 )+ (\delta_L + \delta_R ) (f_2^\dagger f_1 + f_1^\dagger f_2 )+ (\delta_L - \delta_R ) (f^\dagger_2 f^\dagger_1 + f_1 f_2),
\end{eqnarray}

and
\begin{eqnarray}
\mathcal{H}_{T} = T( f_1^\dagger d^\dagger + d f_1 + f_1^\dagger d + d^\dagger f_1).
\end{eqnarray}

Using the fermionic states, a low-energy state of the system can be described by
\begin{eqnarray}
|G \rangle = P_1 |0\rangle  +  P_2   f_1^{\dagger}   d^{\dagger} |0\rangle
+ P_3 f_2^{\dagger} f_1^{\dagger}  |0\rangle+  P_4 f_2^{\dagger} d^{\dagger}  |0\rangle
+ P_5 f_1^{\dagger}  |0\rangle+  P_6 d^{\dagger}  |0\rangle+ P_7 f_2^{\dagger}|0\rangle  + P_8  f_2^{\dagger} f_1^{\dagger}   d^{\dagger} |0\rangle,
\end{eqnarray}
and the Hamiltonian is recasted into a 8x8 matrix

\begin{eqnarray}
  \mathcal{H}  &&=
  \left(\begin{array}{cccccccc}
- {J}\cos \frac{\phi}{2} & T &\delta_L-\delta_R&0                            &0&0&0&0\\
T   &  {J}\cos \frac{\phi}{2} +\epsilon &0&\delta_L+\delta_R                  &0&0&0&0\\
\delta_L-\delta_R&0 & {J}\cos \frac{\phi}{2}& T                              &0&0&0&0\\
0&\delta_L-\delta_R& T &  -{J}\cos \frac{\phi}{2}+\epsilon                   &0&0&0&0\\
0&0&0&0&                  {J}\cos \frac{\phi}{2} & T & \delta_L+\delta_R & 0\\
0&0&0&0&                  T &  -{J}\cos \frac{\phi}{2}+\epsilon & 0 & \delta_L-\delta_R\\
0&0&0&0&                   \delta_L+\delta_R & 0 & -{J}\cos\frac{\phi}{2} & T \\
0&0&0&0&                    0 & \delta_L-\delta_R & T & {J}\cos\frac{\phi}{2}+\epsilon
   \end{array}\right).
\end{eqnarray}

It is block-diagonalized into two 4x4 matrices, for even- and odd-paritiy subspaces respectively.

\begin{center}
\noindent\textbf{B. Floquet theory}\\
\end{center}

The Hamiltonian $\mathcal{H}(t)$ in Eq.~(3) in main text is periodic with a period of $T=2\pi/\Omega$. Its Fourier transformation is given by
\begin{eqnarray}
\mathcal{H}(t) = \sum_m \mathcal{H}_{m} e^{i m \Omega t},
\end{eqnarray}
with

 \begin{eqnarray}
{\mathcal{H}}_0 =  \left(\begin{array}{cccc}
 -{J}\cos \frac{\phi}{2}& T_0                 &\delta_{\rm L}-\delta_{\rm R}&0\\
 T_0  &  {J}\cos \frac{\phi}{2}+\epsilon     &0&\delta_{\rm L}+\delta_{\rm R}\\
\delta_{\rm L}-\delta_{\rm R}&0&               {J}\cos \frac{\phi}{2} &T_0 \\
0&\delta_{\rm L}+\delta_{\rm R}&       T_0 &-{J}\cos\frac{\phi}{2} +\epsilon \\
   \end{array}\right),
{\mathcal{H}}_1 = {\mathcal{H}}_{-1}=  \left(\begin{array}{cccc}
 0& T_1 &0&0\\
 T_1 & 0&0&0\\
0&0& 0 & T_1  \\
0&0& T_1&0\\
   \end{array}\right),
\end{eqnarray}

with all other components zero.

\end{widetext}

Floquet theorem, a time version of Bloch theorem in crystal with periodic potential in space\cite{mahan}, states that the wave functions governed by this time-periodic Hamiltonian is given by
\begin{eqnarray}
\Psi(t) \equiv [P_1(t), P_2(t), P_3(t), P_4(t)]^{\rm t} =  \psi(t) e^{-iQ t},
\end{eqnarray}
with $\psi(t) = \psi(t+2\pi/\Omega)$ where $Q$ is to be determined.
Now we perform Fourier transform on the wave function with period $T=2\pi/\Omega$,
\begin{eqnarray}
 \psi (t) = \sum_n \psi_{n} e^{i n \Omega t},
\end{eqnarray}
where $\psi_n$ is a vector with 4 components $\psi_{n,\alpha}$.

We look back to Eq.~(3) in main text which can be written in a compact form
\begin{eqnarray}
 && {i  } \frac {d} {dt} \Psi (t) = \mathcal{H}(t) \Psi(t),
\end{eqnarray}
where $\hbar=1$ is understood.
Plugging the Fourier transformations into the above equation, it is reformulated into
\begin{eqnarray}
 \sum_n(Q-n\Omega) \psi_{n,\alpha} e^{i n \Omega t}  = \sum_{m,l,\beta} \mathcal{H}_{m,\alpha,\beta} \psi_{l,\beta} e^{i(m+ l )\Omega t},\nonumber\\
\end{eqnarray}
where $\alpha,\beta$ indicate the elements of each 4x4 matrix $\mathcal{H}_m$. It is clear that one should have
\begin{eqnarray}
Q \psi_{n,\alpha}   = \sum_{l,\beta}\left( \mathcal{H}_{n-l,\alpha,\beta} + n \Omega I_{\alpha,\beta} \right)\psi_{l,\beta},
\end{eqnarray}
where $I_{\alpha,\beta}$ is the element of the 4x4 unit matrix.
We now have a static eigen-problem, with eigen value $\Omega$ and eigen function $\psi_{n,\alpha}$. Defining the eigen vector $ \tilde\psi = (..., \psi^{\rm t}_{-1}, \psi^{\rm t}_0, \psi^{\rm t}_1,...)^{\rm t}$, the resultant Floquet Hamiltonian should be
\begin{eqnarray}
&&{\mathcal{H}}_{\rm f} = \\\nonumber
&&\left(\begin{array}{ccccccc}
 \cdot&\cdot&\cdot&\cdot&\cdot&\cdot&\cdot\\
\cdot& {\mathcal{H}}_0-2\Omega \hat I & {\mathcal{H}}_1&0& 0&0& \cdot\\
\cdot& {\mathcal{H}}_1 & {\mathcal{H}}_0-\Omega \hat I&{\mathcal{H}}_1& 0&0& \cdot\\
\cdot& 0 & {\mathcal{H}}_1& {\mathcal{H}}_0& {\mathcal{H}}_1&0& \cdot\\
\cdot& 0 & 0&{\mathcal{H}}_1& {\mathcal{H}}_0+\Omega \hat I&{\mathcal{H}}_1& \cdot\\
\cdot& 0 & 0&0& {\mathcal{H}}_1&{\mathcal{H}}_0+2\Omega \hat I& \cdot\\
 \cdot&\cdot&\cdot&\cdot&\cdot&\cdot&\cdot\\
   \end{array}\right).\nonumber
\end{eqnarray}

The problem of solving time-dependent Schr\"{o}ding equation (32) now is reduced to a static eigen problem. The price one has to pay is to expand a 4x4 matrix into an matrix with infinite dimensions. Now we show how the Floquet Hamiltonian determines the quantum evolution of the system. We define a Floquet state $|n,\alpha\rangle$ by $\psi_{n,\alpha}$ with $\alpha$ denoting one of the quantum states $|S_1\rangle, |S_2\rangle, |S_3\rangle$ and $|S_4\rangle$ and $n$ the Fourier component.
We notice that the eigen-value of the Floquet Hamiltonian has a periodic property, that is,
if $Q$ is a eigen value, then $Q + n\Omega$ is also an eigen value. This periodic property comes from the block-diagonal nature of the Floquet hamiltonian. Therefore,  we can label eigen values by,
\begin{eqnarray}
Q_{n,\gamma} = q_\gamma+ n\Omega,
\end{eqnarray}
where
\begin{eqnarray}
q_\gamma= Q_{0,\gamma}
\end{eqnarray}
denote the four eigen values with smallest absolute values. Meanwhile, the eigen functions are also periodic, that is, for any eigen vector $|Q_{n,\gamma}\rangle$ corresponding to eigenvalue $Q_{n,\gamma}$, we have
\begin{eqnarray}
\langle l, \alpha|Q_{n,\gamma}\rangle = \langle l+m , \alpha|Q_{n+m,\gamma}\rangle.
\end{eqnarray}
It is clear that despite the infinite dimensionless of the Floquet Hamiltonian $\mathcal{H}_{\rm f}$, there are actually only four independent eigen-values and eigen-vectors. We can extract all information and reconstruct the time-dependent wave-function with the four eigen values $q_\gamma$ and the four corresponding eigen-vectors $|q_\gamma \rangle$.

Now we consider the transition probability between two states $|\alpha \rangle$ and $|\beta\rangle$ as a function of time. The evolution operator between these two states is given by\cite{shirley}
\begin{eqnarray}
\hat U_{\beta,\alpha} (t;t_0) =  \sum_{n,\gamma} P_\gamma \psi_{n,\beta,\gamma}(t),
\end{eqnarray}
with $P_\gamma = \sum_{n} \psi^*_{n,\alpha,\gamma} e^{-in\Omega t_0} e^{i Q_\gamma t_0}$ provided that the system starts from the state $|\alpha \rangle$ at time $t_0$.
Taking advantage of Eq.~(38), one can prove
\begin{eqnarray}
\hat U_{\beta,\alpha} (t;t_0) =  \sum_n \langle n,\beta|e^{-i\mathcal{ H}_{\rm f} (t-t_0)}|0,\alpha\rangle e^{in\Omega t}.
\end{eqnarray}
Considering that the transition probability should be an average of $t_0$ with $t-t_0$ constant, since the initial phase of the system is not well defined, we obtain the transition probability
\begin{eqnarray}
P_{\alpha->\beta} = \sum_{n} \left|\langle n,\beta|e^{-i\hat H_f (t-t_0)}|0,\alpha\rangle \right|^2.
\end{eqnarray}
Now the transition probability between the two states governed by the time-periodic Hamiltonian is given in terms of the summation on transition probabilities between Floquet states governed by time-independent
Floquet Hamiltonian. For short-range couplings between blocks associated with different photons in Floquet Hamiltonian,
namely for ac driving with a few Fourier components, the number of terms in the above summation is limited.

\begin{center}
\noindent\textbf{C. Rabi oscillations}\\
\end{center}

 Next we analyze possible Rabi oscillatons in the system. In a resonant oscillation, two Floquet states become degenerate in energy, which can be captured by a 2x2 matrix for these two states, while all other matrix elements are taken into account as perturbations. These perturbations slightly shift the energies of the two Floquet states, and modulate the matrix element between them as well. The effective 2x2 matrix will provide the oscillation conditions.

Let us first discuss the oscillation between $|S_1\rangle$ and $|S_2\rangle$. There is a contribution from the Floquet states $|0, S_1\rangle$ and $|-1,S_2\rangle$ with the bare 2x2 matrix
\begin{eqnarray}
\mathcal{H} = \left(\begin{array}{cc}
  -{J}\cos \frac{\phi}{2}  &T_1 \\
    T_1   &  {J}\cos \frac{\phi}{2}  +\epsilon -\Omega
   \end{array}\right).
\end{eqnarray}
It is obvious that there is a resonant Rabi oscillation around $\Omega \approx 2J \cos \frac{\phi}{2} +\epsilon$.
We now consider contributions from other states in terms of the second-order perturbations. Since $|0,S1\rangle$ has matrix elements with $|0,S_2\rangle$, $|0,S_3\rangle$, $|1,S_2\rangle$, its energy is shifted by
\begin{eqnarray}
E_{-} &&= - \frac{T_0^2}{2J \cos \frac{\phi}{2}  +\epsilon} - \frac{(\delta_L-\delta_R)^2}{2J \cos \frac{\phi}{2}} -\frac{T_1^2}{4J \cos \frac{\phi}{2}  +2\epsilon}\nonumber\\
&&=- \frac{2T_0^2+T_1^2}{4J \cos \frac{\phi}{2}  +2\epsilon} - \frac{(\delta_L-\delta_R)^2}{2J \cos \frac{\phi}{2}}.
\end{eqnarray}
Similarly, since $|-1,S2\rangle$ has matrix elements with $|-1,S_1\rangle$, $|-1,S_4\rangle$, $|-2,S_1\rangle$, its energy shift is given by
\begin{eqnarray}
E_{+}
= \frac{2T_0^2+T_1^2}{4J \cos \frac{\phi}{2}  +2\epsilon} + \frac{(\delta_L+\delta_R)^2}{2J \cos \frac{\phi}{2}}.
\end{eqnarray}
Noticing that there is no two-step virtual hopping process between these two states, the off-diagonal elements are not modified within the second-order perturbation.
In this way we arrive at the effective Hamiltonian Eq.~(10) in main text. Considering the transformation between $|0,S_1\rangle$ and $|1,S_2\rangle$ in the same way, the condition for the Rabi oscillation
between $|S_1\rangle$ and $|S_2\rangle$ is given by Eq.~(12) in main text.
Rabi oscillation between $|S_3\rangle$ and $|S_4\rangle$ can be derived similarly.

Rabi oscillation takes place between $|S_1\rangle$ and $|S_4\rangle$, for which we should look at the Floquet states $|0,S_1\rangle$ and $|-1,S_4\rangle$.
The bare 2x2 matrix is
\begin{eqnarray}
\mathcal{H} = \left(\begin{array}{cc}
  -{J}\cos \frac{\phi}{2}  &0 \\
    0   &  -{J}\cos \frac{\phi}{2}  +\epsilon -\Omega
   \end{array}\right).
\end{eqnarray}
These two states are connected with each other through $|-1,S_2\rangle $  and $|0,S_3\rangle $, which produce nonzero off-diagonal elements through the second-order perturbation,
\begin{eqnarray}
\mathcal{H}_{14} &&= \sum_{n,\alpha} \frac{  \langle0,S1|  \hat H  |n,\alpha\rangle  \langle n,\alpha| \hat H |-1,S4\rangle  }{ \langle0,S1|  \hat H |0,S1\rangle - \langle n,\alpha|  \hat H |n,\alpha\rangle }\nonumber\\\
&&= \frac{ T_1(\delta_L-\delta_R)  }{-{J}\cos \frac{\phi}{2} - {J}\cos \frac{\phi}{2}} + \frac{ (\delta_L+\delta_R) T_1  }{ -{J}\cos \frac{\phi}{2}-{J}\cos \frac{\phi}{2} }\nonumber\\\
&&=-\frac{T_1 \delta_L}{J\cos \frac{\phi}{2}}.
\end{eqnarray}
The energy shift for state $|0,S_1\rangle$ is

\begin{widetext}
\begin{eqnarray}
E'_{-} = - \frac{T_0^2}{2J \cos \frac{\phi}{2}  +\epsilon} - \frac{(\delta_L-\delta_R)^2}{2J \cos \frac{\phi}{2}} -\frac{T_1^2}{2J \cos \frac{\phi}{2}  + 2\epsilon}-\frac{T_1^2}{2J \cos \frac{\phi}{2}},
\end{eqnarray}
and for $|-1,S_4\rangle$
\begin{eqnarray}
E'_{+} =  \frac{T_0^2}{2J \cos \frac{\phi}{2}  +\epsilon} + \frac{(\delta_L+\delta_R)^2}{2J \cos \frac{\phi}{2}} + \frac{T_1^2}{2J \cos \frac{\phi}{2}  + 2\epsilon}+\frac{T_1^2}{2J \cos \frac{\phi}{2}}.
\end{eqnarray}
\end{widetext}

We then arrive at the effective 2x2 matrix within the second-order perturbation
\begin{eqnarray}
\mathcal{H} = \left(\begin{array}{cc}
  -{J}\cos \frac{\phi}{2} +E'_-  & \frac{T_1 \delta_L}{{J}\cos{\phi}/{2}}\\
   \frac{T_1 \delta_L}{{J}\cos {\phi}/{2}}   &  -{J}\cos \frac{\phi}{2}  +\epsilon -\Omega +E'_+
   \end{array}\right).\nonumber\\
\end{eqnarray}
A Rabi oscillation appears at $\Omega= \epsilon +E'_+ - E'_-$, with frequency of $T_1 \delta_L/(J\cos\frac{\phi}{2})$.

We notice two interesting properties of this Rabi oscillation. First, its frequency is smaller than the others discussed above by order of $\delta_L/J$, since this oscillation is driven by second-order hopping processes.
Second, this frequency contains only $\delta_L$, while $\delta_R$ is missing. This is related to the structure of setup where QD is connected to junction MBS on the right segment (see Fig.~1 of main text).
The transformation from $|S_1\rangle$ to $|S_4\rangle$ is contributed from four possible processes. For the first two processes, a) an electron jumps from the junction to QD and then another electron jumps from the left segment to junction
\begin{eqnarray}
(\delta_L \gamma_1 \gamma_3) (iT d^\dagger \gamma_2),
\end{eqnarray}
and b) an electron jumps from the left segment to junction and then jumps to QD
\begin{eqnarray}
 (iT d^\dagger \gamma_2)(\delta_L \gamma_1 \gamma_3).
\end{eqnarray}
The total contribution is $2iT\delta_Ld^\dagger \gamma_2 \gamma_1 \gamma_3$.
Similarly, the other two processes include an electron coming from right segment, which are described by
 \begin{eqnarray}
(\delta_R \gamma_4 \gamma_2) (iT d^\dagger \gamma_2),
\end{eqnarray}
and
  \begin{eqnarray}
(iT d^\dagger \gamma_2)(\delta_R \gamma_4 \gamma_2).
\end{eqnarray}
However, in contrast to the previous case, the two processes interfere with each other destructively
  \begin{eqnarray}
iT \delta_R (d^\dagger \gamma_2  \gamma_4 \gamma_2 + \gamma_4 \gamma_2 d^\dagger \gamma_2 ) = 0,
\end{eqnarray}
and make a vanishing contribution to the transformation. In other words, the oscillation between $|S_1\rangle$ to $|S_4\rangle$ only involves electron hopping from the left segment, thus $\delta_R$ is missing in the Rabi frequency. One can understand the Rabi oscillation between $|S_2\rangle$ to $|S_3\rangle$ in the same way.

Finally, we should note that the above discussions in terms of 2x2 matrix are based on perturbation approach, and therefore are not valid when the energy shift and off-diagonal elements are large. Especially, when $\cos \phi /2 =0$ or $2J \cos \phi/2 \pm \epsilon =0$, denominators in energy shifts and coupling elements become zero. These cases correspond to an additional energy degeneracy besides the two states under concern, where two sets of Rabi oscillatons take place simultaneously as can be seen in Fig.~3 of main text.

\begin{center}
\noindent\textbf{D. Multi-state oscillation}\\
\end{center}

Here we analyze the multi-state oscillation shown in the Fig.~4 of main text at $\Phi= \Phi_0 /2$ and $\Omega=\epsilon$ within the Floquet formalism.
At this special point, the four Floquet states
$|0,S_1\rangle$, $|-1,S_2\rangle$, $|0,S_3\rangle$, and $|-1,S_4\rangle$ with approximately zero energy are involved into resonant oscillation.
The un-perturbed Hamiltonian is given by
 \begin{eqnarray}
{\mathcal{H}} =  \left(\begin{array}{cccc}
 0 & T_1                 &\delta_{\rm L}-\delta_{\rm R}&0\\
 T_1  &  0     &0&\delta_{\rm L}+\delta_{\rm R}\\
\delta_{\rm L}-\delta_{\rm R}&0&               0 &T_1 \\
0&\delta_{\rm L}+\delta_{\rm R}&       T_1 &0 \\
   \end{array}\right),\nonumber\\
\end{eqnarray}
Interchanging the order of the basis state $|-1,S_2\rangle$ and $|0,S_3\rangle$, the matrix is transformed to,
 \begin{eqnarray}
{\mathcal{H}} =  \left(\begin{array}{cccc}
 0 & \delta_{\rm L}-\delta_{\rm R} &T_1                &0\\
\delta_{\rm L}-\delta_{\rm R}  &  0     &0& T_1\\
T_1&0&               0 &\delta_{\rm L}+\delta_{\rm R} \\
0&T_1&       \delta_{\rm L}+\delta_{\rm R}&0 \\
   \end{array}\right).
\end{eqnarray}
With the "bonding" states
 \begin{eqnarray}
 &&|S^+_{13}\rangle \equiv \frac{1}{\sqrt{2}}(|0,S_1\rangle+|0,S_3\rangle), \nonumber\\
 &&|S^+_{24}\rangle \equiv \frac{1}{\sqrt{2}}(|-1,S_2\rangle+|-1,S_4\rangle),
 \end{eqnarray}
and "anti-bonding" states
 \begin{eqnarray}
 &&|S^-_{13}\rangle \equiv \frac{1}{\sqrt{2}}(|0,S_1\rangle-|0,S_3\rangle), \nonumber\\
 &&|S^-_{24}\rangle \equiv \frac{1}{\sqrt{2}}(|-1,S_2\rangle-|-1,S_4\rangle),
 \end{eqnarray}
the Hamiltonian is transformed into
 \begin{eqnarray}
{\mathcal{H}} =  \left(\begin{array}{cccc}
\delta_{\rm L}-\delta_{\rm R} & 0 &T_1                &0\\
0  &  -\delta_{\rm L}+\delta_{\rm R}     &0& T_1\\
T_1&0&               \delta_{\rm L}+\delta_{\rm R} & 0 \\
0&T_1&       0 &-\delta_{\rm L}-\delta_{\rm R} \\
   \end{array}\right)
\end{eqnarray}
in the new basis $|S^+_{13}\rangle$, $|S^-_{13}\rangle$, $|S^+_{24}\rangle$, $|S^-_{24}\rangle$.
For example, the matrix elements in the first row are calculated as,

\begin{widetext}
 \begin{eqnarray}
\langle S^+_{13}| {\mathcal{H}} |S^+_{13}\rangle &&=  \frac{1}{{2}} \left[\langle S_{1}| {\mathcal{H}} |S_{1}\rangle + \langle S_{1}| {\mathcal{H}} |S_{3}\rangle + \langle S_{3}| {\mathcal{H}} |S_{1}\rangle + \langle S_{3}| {\mathcal{H}} |S_{3}\rangle\right] \nonumber\\
&&= \frac{1}{{2}} \left[ 0 +(\delta_L - \delta_R) + (\delta_L - \delta_R) + 0 \right] =\delta_L -\delta_R,
 \end{eqnarray}

 \begin{eqnarray}
\langle S^+_{13}| {\mathcal{H}} |S^-_{13}\rangle
&&=  \frac{1}{{2}} \left[\langle S_{1}| {\mathcal{H}} |S_{1}\rangle
- \langle S_{1}| {\mathcal{H}} |S_{3}\rangle
+ \langle S_{3}| {\mathcal{H}} |S_{1}\rangle
- \langle S_{3}| {\mathcal{H}} |S_{3}\rangle\right] \nonumber\\
&&= \frac{1}{{2}} \left[ 0 -(\delta_L - \delta_R) + (\delta_L - \delta_R) + 0 \right] =0,
 \end{eqnarray}

 \begin{eqnarray}
\langle S^+_{13}| {\mathcal{H}} |S^+_{24}\rangle
&&=  \frac{1}{{2}} \left[\langle S_{1}| {\mathcal{H}} |S_{2}\rangle
+ \langle S_{1}| {\mathcal{H}} |S_{4}\rangle
+ \langle S_{3}| {\mathcal{H}} |S_{2}\rangle
+ \langle S_{3}| {\mathcal{H}} |S_{4}\rangle\right] \nonumber\\
&&= \frac{1}{{2}} \left[ T_1 +0 +0 + T_1 \right] = T_1,
 \end{eqnarray}

 \begin{eqnarray}
\langle S^+_{13}| {\mathcal{H}} |S^-_{24}\rangle
&&=  \frac{1}{{2}} \left[\langle S_{1}| {\mathcal{H}} |S_{2}\rangle
- \langle S_{1}| {\mathcal{H}} |S_{4}\rangle
+ \langle S_{3}| {\mathcal{H}} |S_{2}\rangle
- \langle S_{3}| {\mathcal{H}} |S_{4}\rangle\right] \nonumber\\
&&= \frac{1}{{2}} \left[ T_1 +0 +0 - T_1 \right] = 0,
 \end{eqnarray}
 \end{widetext}

and all other elements can be obtained similarly.
Now the Hamiltonian can be block diagonalized by reordering the basis as $|S^+_{13}\rangle$, $|S^+_{24}\rangle$, $|S^-_{13}\rangle$, $|S^-_{24}\rangle$,
 \begin{eqnarray}
{\mathcal{H}} =  \left(\begin{array}{cccc}
\delta_{\rm L}-\delta_{\rm R} & T_1 & 0                &0\\
T_1  &  \delta_{\rm L}+\delta_{\rm R}     &0& 0\\
0&0&               -\delta_{\rm L}+\delta_{\rm R} & T_1 \\
0&0&       T_1 &-\delta_{\rm L}-\delta_{\rm R} \\
   \end{array}\right).\nonumber\\
\end{eqnarray}

The physics can be understood in a more transparent way. For $\Phi= \Phi_0 /2$, there is no coupling among the two MBSs $\gamma_1$ and $\gamma_2$. It is
then better to describe the system by parity state of electron number on the left segment formed by $\gamma_1$ and $\gamma_3$, and that on the right segment
formed by $\gamma_4$ and $\gamma_2$. It is not difficult to see that the bonding state $|S^+_{13}\rangle$ corresponds to even parity states on both left
and right segments, while $|S^+_{24}\rangle$ corresponds to even parity on the left segment while odd parity on the right segment, and the two states
are coupled via $T_1$ which changes the parity of the right segment using QD. The same applies for the two anti-bonding states. The absence of direct coupling
between the two blocks is due to the structure of device setup where QD is only connected to the right segment.

Now we consider the energy shifts for these 'bonding' and 'anti-bonding' states with the second-order perturbation treatment.
Since $\epsilon\gg T_0\gg T_1, \delta$, dominant
contributions should come from virtual processes from Floquet states in the same block. For $|S^+_{13}\rangle$,  the two Floquet states $|0,S_2\rangle$ and $|0,S_4\rangle$ with the same energy $\epsilon$ and same coupling matrix element $T_0$ with $|S^+_{13}\rangle$ contribute an energy shift
 \begin{eqnarray}
E^{''}_1 = \frac{2T^2_0}{\delta_{\rm L}-\delta_{\rm R} - \epsilon}.
\end{eqnarray}
For $|S^+_{24}\rangle$, the two Floquet states $|-1,S_1\rangle$ and $|-1,S_3\rangle$ with energy $-\epsilon$ and coupling matrix element $T_0$ with $|S^+_{24}\rangle$ contribute an energy shift
 \begin{eqnarray}
E^{''}_2 = \frac{2T^2_0}{\delta_{\rm L}+\delta_{\rm R} + \epsilon}.
\end{eqnarray}

Similarly, the anti-bonding state $|S^-_{13}\rangle$ is connected to $|0,S_2\rangle$ and $|0,S_4\rangle$ with coupling matrix elements $T_0$ and $-T_0$ respectively, getting an energy shift
 \begin{eqnarray}
E^"_3  = \frac{2T^2_0}{-\delta_{\rm L}+\delta_{\rm R} - \epsilon},
\end{eqnarray}
and $|S^-_{24}\rangle$ is connected to $|-1,S_1\rangle$ and $|-1,S_3\rangle$ with coupling matrix elements $T_0$ and $-T_0$ respectively, getting an energy shift
 \begin{eqnarray}
E^"_4  = \frac{2T^2_0}{-\delta_{\rm L}-\delta_{\rm R} + \epsilon}.
\end{eqnarray}

The matrix up to the second-order perturbation is given as

 \begin{widetext}
 \begin{eqnarray}
{\mathcal{H}} =  \left(\begin{array}{cccc}
\delta_{\rm L}-\delta_{\rm R} + E^"_1 & T_1 & 0                &0\\
T_1  &  \delta_{\rm L}+\delta_{\rm R} + E^"_2     &0& 0\\
 0&0&               -\delta_{\rm L}+\delta_{\rm R} + E^"_3 & T_1 \\
0&0&       T_1 &-\delta_{\rm L}-\delta_{\rm R}  + E^"_4 \\
   \end{array}\right).
\end{eqnarray}
\end{widetext}

There is a Rabi oscillation between the two bonding states with frequency
\begin{eqnarray}
\omega_1 =  \sqrt{(2\delta_R + E^"_2 - E^"_1)^2 + T_1^2},
\end{eqnarray}
and between the two anti-bonding states with frequency
\begin{eqnarray}
\omega_2 =  \sqrt{(2\delta_R + E^"_3 - E^"_4)^2 + T_1^2}.
\end{eqnarray}
It is noticed that the oscillations are not full (see Eq.~(11) in main text), and thus the two frequencies are
not equal even though the off-diagonal coupling elements in the two blocks are both $T_1$.

\begin{center}
\noindent\textbf{E. Results based on numerical simulation}\\
\end{center}

\begin{figure}
\begin{center}
\includegraphics[clip=true,width=0.8\columnwidth]{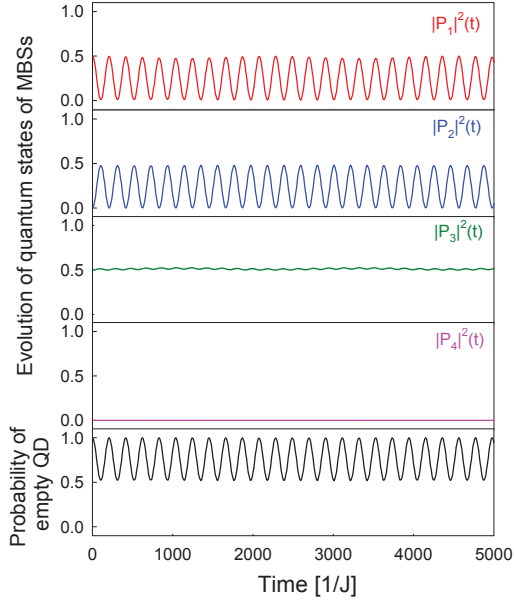}
\caption{(Color online). Time evolution of the quantum state of MBSs
and the QD occupation probability with initial condition of $|P_1|^2=|P_3|^2=1/2$ for $\Phi=0.1\Phi_0$ and $\Omega=2.3J$, and other parameters are the same as in Fig.~2 in the main text except for $T_1=0.015J$. $\hbar=1$ is understood through the Supplementary Material. }
\end{center}
\end{figure}

  \begin{figure}
\begin{center}
\includegraphics[clip=true,width=0.8\columnwidth]{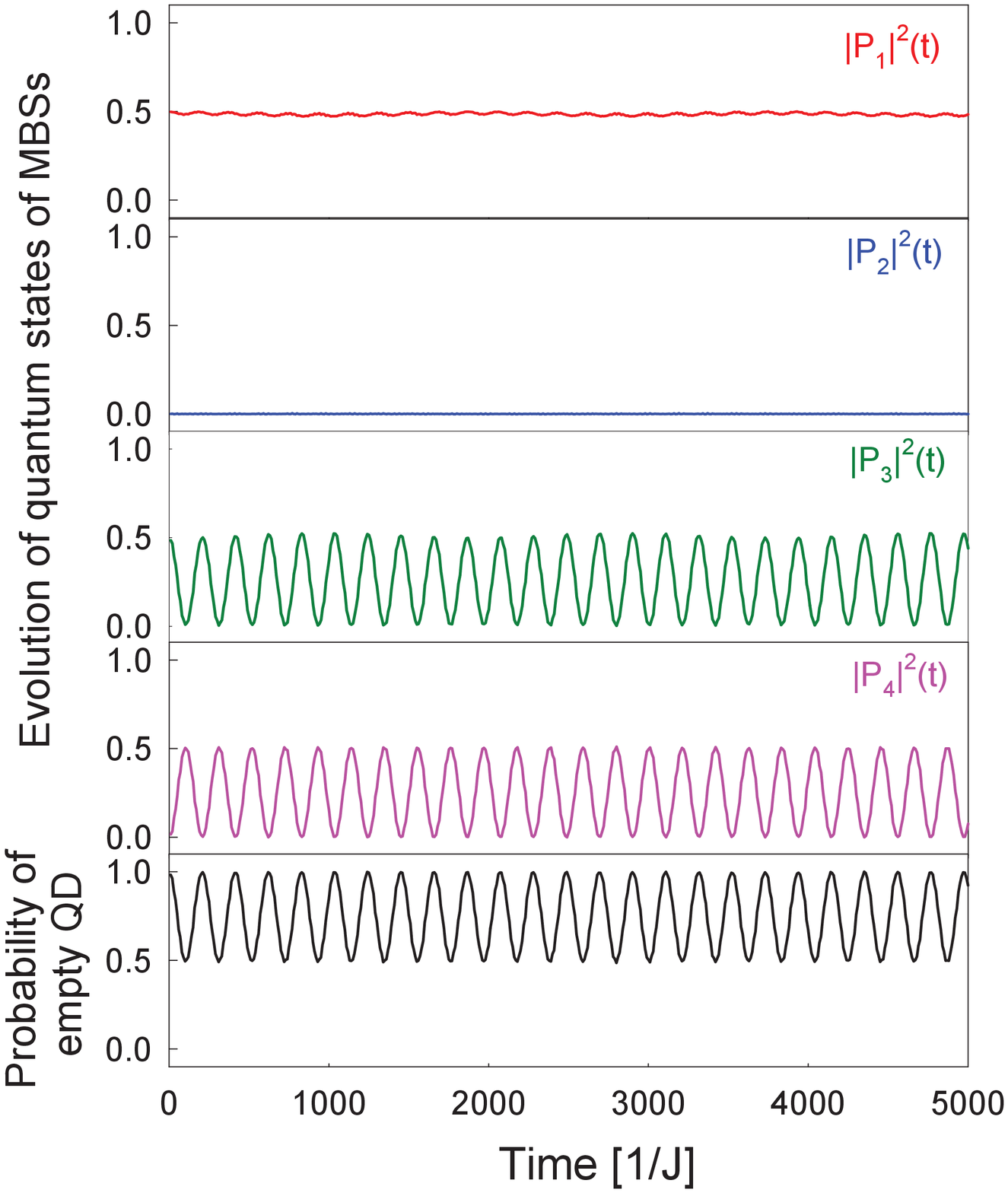}
\caption{(Color online). Same as Fig.~5 except for $\Omega=1.5J$. }
\end{center}
\end{figure}

\begin{figure}
\begin{center}
\includegraphics[clip=true,width=0.8\columnwidth]{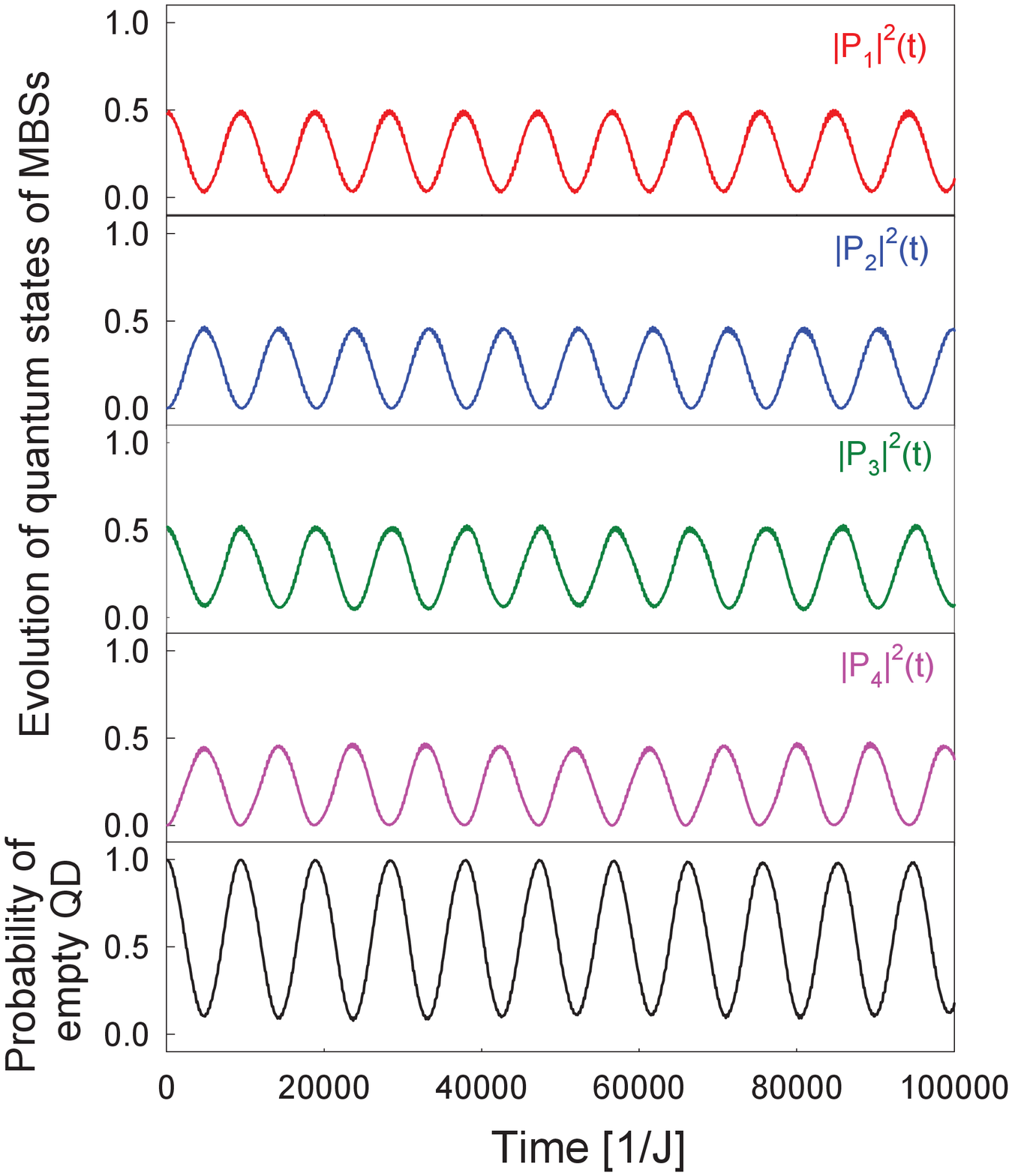}
\caption{(Color online). Same as Fig.~5 except for $\Omega=0.4J$.}
\end{center}
\end{figure}

\begin{figure}
\begin{center}
\includegraphics[clip=true,width=0.8\columnwidth]{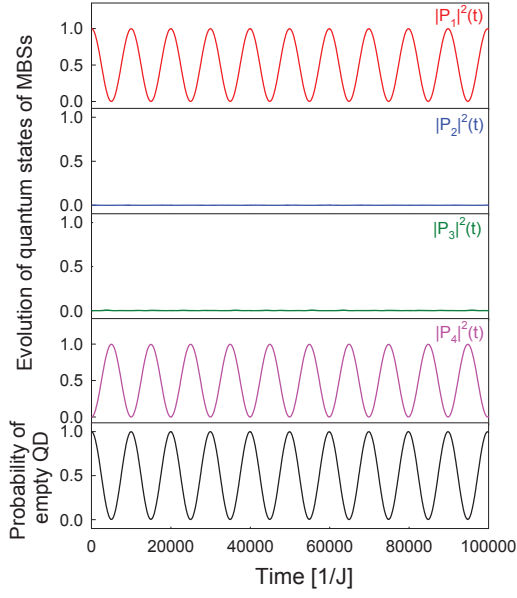}
\caption{(Color online). Same as Fig.~7 except for that the initial state is taken as $P_1=1$.}
\end{center}
\end{figure}

\begin{figure}
\begin{center}
\includegraphics[clip=true,width=0.8\columnwidth]{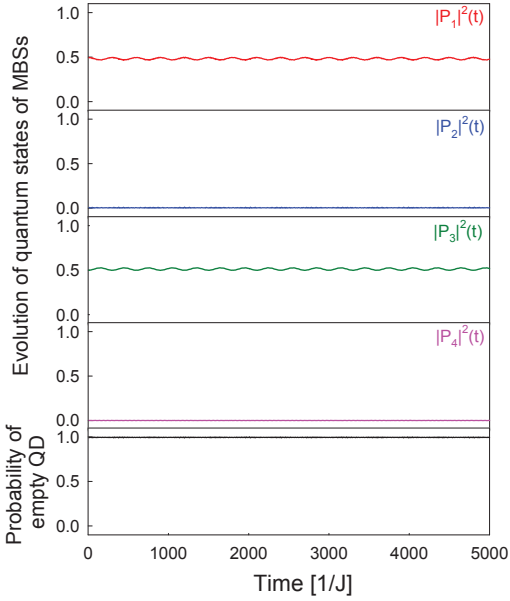}
\caption{(Color online). Same as Fig.~5, with $\Omega=2J$.}
\end{center}
\end{figure}

\begin{figure}
\begin{center}
\includegraphics[clip=true,width=0.8\columnwidth]{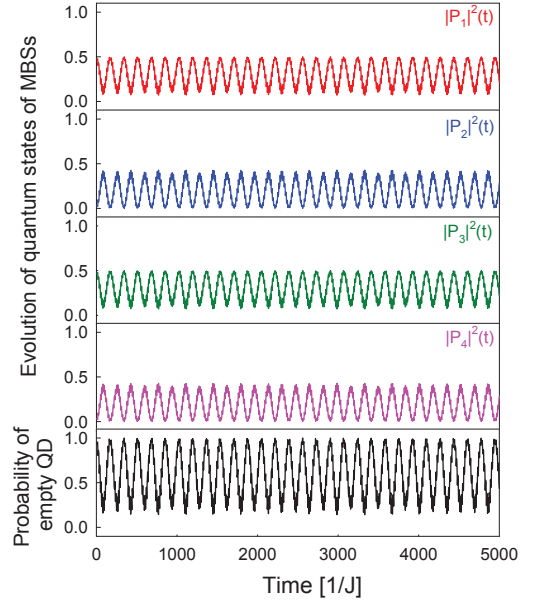}
\caption{(Color online). Same as Fig.~5 except for $\Phi=\Phi_0/2$ and $\Omega=\epsilon=0.4J$, the crossing point of two curves and the horizontal line in Fig.7 of main text.}
\end{center}
\end{figure}

\begin{figure}[t]
\begin{center}
\includegraphics[clip=true,width=0.8\columnwidth]{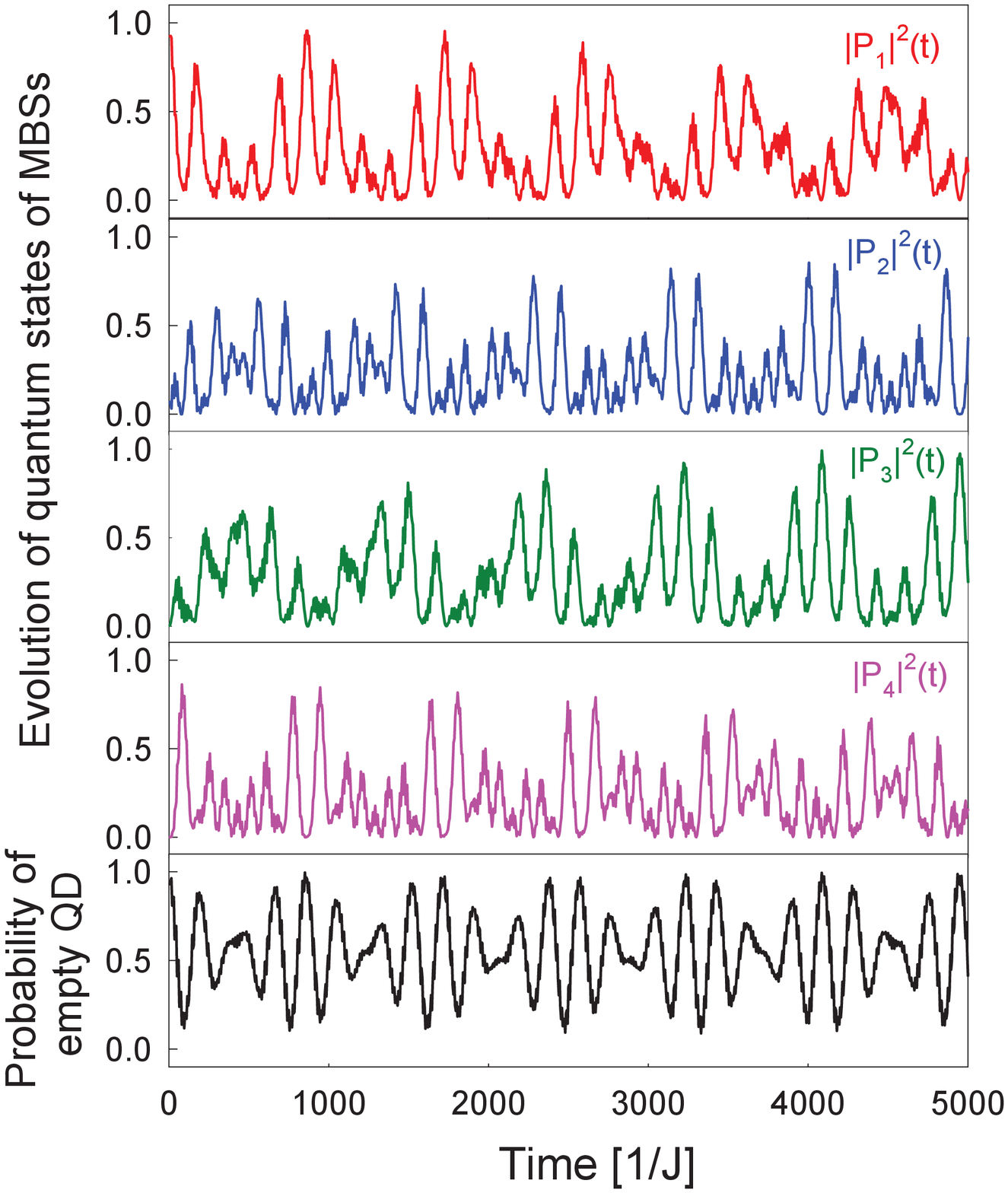}
\caption{(Color online). Same as Fig.~10 except for the initial state is taken as $P_1=1$.}
\end{center}
\end{figure}

Besides the analytic treatment based on Floquet theorem, we also perform numerical calculation by integrating directly the time-dependent Schr\"{o}dinger equation (3) in main text. The results are summarized in Fig.~3 of main text, which are in good agreement with analytic results, except for the region around the crossing points of curves where the simple treatment of 2x2 matrix breaks down.
In numerical simulations, we start from a quantum state with empty QD with $|P_1|^2=|P_3|^2=1/2$. The strength of Rabi oscillation given by color scale in Fig.~3 in main text is measured by the difference in occupation probability of electron on QD during the Rabi oscillation.

In order to understand the results summarized in Fig.~3 in main text, we show the detailed Rabi oscillations at typical parameter points.
In Fig.~5, we plot the evolution of quantum states and QD occupation state for $\Phi= 0.1\Phi_0$ and $\Omega=2.3J$ which is on the upper curve in Fig.~3 of main text.
We can see clearly oscillations between states $|S_1\rangle$ and $|S_2\rangle$, and oscillation in QD occupation probability which can be measured in experiments.
In Fig.~6, we show the evolutions for $\Phi= 0.1\Phi_0$ and $\Omega=1.5J$ on the lower curve. In this case oscillations appear between states $|S_3\rangle$ and $|S_4\rangle$, accompanied by oscillation in QD occupation probability. In Fig.~7, we illustrate the evolutions for $\Phi= 0.1\Phi_0$ and $\Omega=0.4J$ on the horizontal line. In this case, oscillation between $|S_1\rangle$ and $|S_4\rangle$ and that between $|S_2\rangle$ and $|S_3\rangle$  take place simultaneously. We particularly note that the period of these oscillations is much longer than those in Figs.~5 and 6. The two independent oscillations can be seen clearly in Fig.~S4 where an initial state with
$|P_1|^2=1$ is taken. For $\Phi= 0.1\Phi_0$ and $\Omega=2J$, there is no oscillation in the system as in Fig.~S5, corresponding to the dark area in Fig.~3 in main text.
At $\Phi= \Phi_0/2$ and $\Omega=0.4J$, there is only one Rabi oscillation with frequency $\omega_1$ as shown in Fig.~S6 if
one starts from the initial state of $|S^+_{13}\rangle =  \frac{1}{\sqrt{2}}(|0, S_{1}\rangle + |0, S_{3}\rangle)$; starting from the initial state
$|0,S_1\rangle = \frac{1}{\sqrt{2}} (|S^+_{13}\rangle + |S^-_{13}\rangle  )$, two Rabi oscillations take place with the two frequencies $\omega_1$ and $\omega_2$, which results in beating waves in QD occupation probability as shown in Fig.~S7 and Fig.~4 in main text.

\end{document}